# Generating Classes of 3D Virtual Mandibles for AR Applications

Felix G. Hamza-Lup, Neha R. Hippalgaonkar, Alexa D. Sider, Anand P. Santhanam,
Gerald Bertetta, Celina Imielinska, and Jannick P. Rolland

**Abstract -** Simulation and modeling represent promising tools for several application domains from engineering to forensic science and medicine. Advances in 3D imaging technology convey paradigms such as Augmented and Mixed Reality (AR/MR) inside promising simulation tools for the training industry. Motivated by the requirement for superimposing anatomically correct 3D models on a Human Patient Simulator (HPS) and visualizing them in an AR environment, the purpose of this research effort is to derive method for scaling a source human mandible to a target human mandible. Results show that, given a distance between two same landmarks on two different mandibles, a relative scaling factor may be computed. Using this scaling factor, results show that a 3D virtual mandible model can be made morphometrically equivalent to a real target-specific mandible within a 1.30 millimeter average error bound. The virtual mandible may be further used as a reference target for registering other anatomical models, such as the lungs, on the HPS. Such registration will be made possible by physical constraints among the mandible and the spinal column in the horizontal normal rest position.

―――――――― ◆ ――――――――

## 1. INTRODUCTION

COMPUTER imaging techniques have become an important aid to diagnosis in the practice of modern medicine. Computed Tomography (CT) and Magnetic Resonance Imaging (MRI) use a sampling or data acquisition process to capture information about the internal anatomy of a living patient. While the information is currently analyzed in 2D, with the advance of 3D displays, Augmented Reality (AR) and Mixed Reality (MR) paradigms, 3D models find an increasing number of applications into the medical field. Key applications include surgical guidance,[1] the development of dynamic anatomy visualization for teaching,[2] and computer-assisted guidance for surgical simulation and training of medical technicians and students [3].

An important requirement for an effective visualization in AR is the use of anatomically accurate 3D virtual models. Specifically, the need for those models within our scope of work is motivated by the requirement for superimposing patient-specific data on a Human Patient Simulator (HPS) and visualizing them in an AR environment for training clinical technicians.

Per our discussions with physicians, an overall coincidence of the 3D large scale models (e.g. lungs) within a 3 mm Root Mean Square (RMS) error with respect to the internal anatomy models would be grandly sufficient for the development of AR training procedures. Such performance imposes an upper bound on the departure of a 3D models from the internal anatomy of the HPS by construction.

Importantly, the mandible is a reference target with anatomical landmarks that can be used for registering anatomical models, such as the lungs and the larynx, on a HPS. Such registration is made possible by physical constraints among the mandible and the spinal column in the horizontal normal rest position as further illustrated in Section 6. While an important limitation in obtaining patient-specific 3D virtual models is the exposure to radiation for CT and the cost of 3D imaging for MRI, the main limitation for training on a HPS is the potential lack of correctly sized models. Thus in the context of training medics on the upper airways of the HPS, the need is to appropriately scale and register a previously acquired 3D mandible from a source human model to fit a target mandible imposed by the HPS. With respect to the goal of registering a scaled given computer generated mandible model to the HPS physical mandible based on common landmarks between the two mandibles, only three non collinear landmarks will be needed given that the mandible is a rigid object. The landmarks selected for registration will not only need to be those most accurately located but also those readily palpable from outside the skin.

The purpose of this study is to demonstrate that, within a gender, race, and age group, it is feasible to scale two 3D high resolution mandibles in order to make them morphometrically equivalent within a 3 mm RMS error which is imposed by the HPS training application. In our study we consider a male Caucasian subject within the age group of 16 to 24 years old where the mandible growth has attained its full maturity, yet it has not started to degenerate. The method used to implement the proper scaling factor will be shown to yield anatomically valid mapping. Specifically in Section 2 of the paper, we review related work on methods used for shape analysis. In Section 3, we describe in detail the virtual mandible model

―――――――――――――

- Felix G. Hamza-Lup is with the Computer Science at the Armstrong Atlantic State University, Savannah,GA,. E-mail: felix@ cs.armstrong.edu.
- Anand P. Santhanam is with the School of Computer Science at the Univ. of Central Florida, Orlando, FL, 32816. E-mail:anand@odalab.ucf.edu
- G. Bertetta is with the College of Health and Public Affairs at the Univ. of Central Florida, Orlando, FL, 32816.
- C. Imielinska is with the Department of Biomedical Informatics at Columbia University
- Jannick P. Rolland is with the College of Optics and Photonics: CREOL &FPCE at the Univ. of Central Florida. E-mail:jannick@odalab.ucf.edu

generation and analysis. Section 4 presents the proposed method of shape analysis for scaling ratio quantification. Section 5 discusses the quantification of the virtual 3D models with respect to their real counterparts. Section 6 provides insights into potential applications in various domains.

## 2. RELATED WORK

Morphometrics generally refers to the analysis of size and shape [4]. While size changes refer to a proportional increase or decrease in all dimensions of a 3D model, shape change refers to a change in the outline of the form under examination [5]. There are two distinct groups of techniques in the current literature: landmark-based techniques and boundary outline techniques. Landmarks are discrete points on a 3D virtual model. Landmark-based techniques are based on establishing distance and angle relations between the landmarks. Boundary outline techniques on the other hand investigate the shape of the perimeter of a structure defined at a certain resolution.

Initial methods of landmark-based analysis were based on either statistical or superimposition approach. In the case of a statistical approach, a suitable 3D model is chosen from a database of 3D models based on the landmark positions(e.g. facial morphing [6]) In the case of a superimposition approach where the 3D models were scaled and rotated until the eigenvectors of the covariance matrix of landmarks of the two 3D models match with each other [7]. Using these methods, the local surface variations were not represented efficiently. Deformation methods such as Finite Element Methods [8] and Thin Plate Splines [9] were used to estimate the shape difference between two 3D virtual models. The required deformation conveyed their morphological differences. Surface reconstruction methods such as elliptical Fourier functions [10] and medial axis methods were also used for boundary outline estimation analysis. These latter methods convey the shape changes between two given 3D virtual models effectively in terms of a set of spherical basis functions. The deformation and surface reconstruction methods provided measurements with high precision but lacked an established relationship to biology and statistical varaitions required for effective analysis [11, 12].

The biological effect on the shapes of anatomical organs is better represented using *Growth Allometry* techniques,[13] where the direction of growth of an anatomical structure based on race and age of a subject is taken into account. An approach to model the mandible bone growth along the surface of the mandible is presented by [14]. Such methods were limited by the lack of mathematical flexibility and user-friendliness.

The most flexible and user-friendly methods for algebraic analysis are the Conventional Cephalometric Methods [15] and analysis based on the landmarks linear distance, angles, and ratios. However these methods do not adequately represent the shape details and are not fully capable of evaluating shape and size. Euclidean Distance Matrix Analysis (EDMA) was proposed as an improvement to these conventional methods [12]. In this method biological shapes are compared using landmark coordinate data by mathematically localizing their morphological distances. The resultant is a set of ratios among the Euclidean distances.

In this paper, we discuss a landmark-based shape analysis in order to scale a given mandible to a target mandible. Such a scaling approach must satisfy the requirement of being able to put the models in coincidence within a 3mm Root Mean Square error. We tailored the EDMA method to the 3D mandible models generation. The distances between landmarks are computed along the surface of the 3D model using geodesic distances.[16-18] The strength of this method is that it combines the mathematical simplicity of the EDMA method discussed in [12] with the biological growth statistics of the mandible discussed in [14]. The work exemplifies translational research driven by a medical application where known methodologies in different areas combined together, as in our case: biomathematics, biology, imaging, modeling, 3D visualization/virtual reality and morphometry are tailored to serve best the application. Translational research in conjunction with well stated medical challenges can serve two causes: solve a sound medical problem and generate potential new open problems in the component areas.

## 3. VIRTUAL 3D MANDIBLE GENERATION

The virtual 3D mandibles were generated via high resolution digitization (i.e. a point is measured within 0.1mm accuracy) as opposed to segmenting CT models because in this first investigation we wanted to ensure the highest possible accuracy of the 3D models.[19, 20] When considering CT, one would want to quantify errors caused by both the limited resolution of the CT acquisition (i.e. $\geq$ 2.5mm) and the segmentation process. From a set of digital 3D points, a 3D polygonal representation of the real object is obtained. The polygonal model can be further processed based on the application needs.

Three Caucasian adult male mandibles shown in Figure 1 were obtained from Global-Technologies [21] .

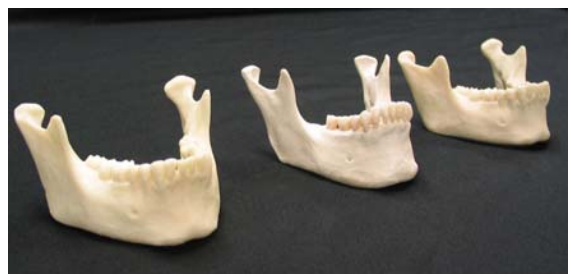

Figure 1. Caucasian adult male mandibles

The computer generated models corresponding to these mandibles were obtained through a non-destructive two-step digitization process performed with an optical tracking system, the Optotrak 3020 [22], and its associated digitizing probe shown in Figure 2a and 2b, respectively

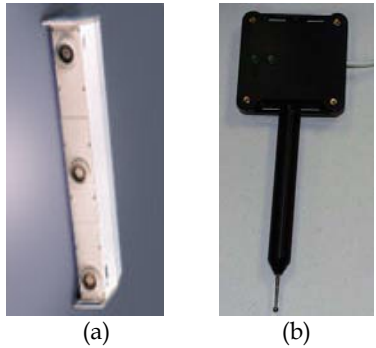

Figure 2. (a) Optotrack 3020 Tracking System; (b) Digitizing Probe

In the first step of digitalization, the position information from the mandible's surface was collected. Naturally the resolution of the generated model is proportional to the number of points collected. For this particular experiment we divided the mandible surface in fifteen regions of different sizes, with larger sizes where the surface curvature was low. Regions slightly overlapped and we collected about 2000 points from each region.

In the second step, the collected data was imported in the *Geomagic Studio 5.0* [23], a software tool that allowed redundant point elimination and generated a 3D polygonal model. The points collected in the fifteen regions were merged for each mandible. The 3D polygonal models obtained consisting of 47k polygons each are illustrated in Figure 3 (a-c).

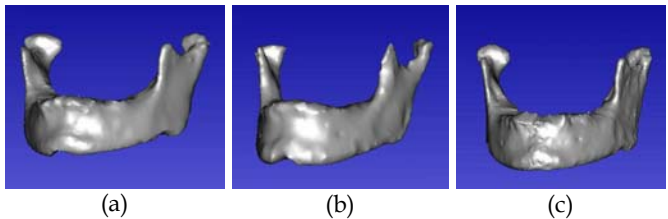

Figure 3. 3D Virtual models of the three mandibles shown in figure 1: (a) Mandible 1; (b) Mandible 2; (c) Mandible 3.

## 4. MANDIBLE SHAPE ANALYSIS

We now describe a method of landmark-based shape analysis. The steps of the method include in section 4.1. the landmark selection, in section 4.2 the shape characterization, and in section 4.3 the scaling computation of one mandible to another.

### 4.1 Landmarks Selection

There are seventy seven (77) significant landmarks on the human mandible [10, 24]. In this experiment thirty-one landmarks were chosen independently by one anatomist and one medical expert based on their biological importance and the direction of growth [14, 25]. The two experts selected the same landmarks as a set with 100% accordance. Specifically they selected 3 landmarks (i.e. number 1, 2, and 3) common to both the left and right sides of a real 3D mandible, and 14 additional landmarks per side, totaling 31 landmarks as shown in Fig.5. Note that the numbering of the landmarks is not the same as that provided in reference [14], rather it follows a 1-31 numbering for the subset of selected landmarks.

The two experts then recorded the position of the 31 selected landmarks on each computer generated mandible in one session. Both experts repeated the procedure once for each mandible at least two weeks later, thus each landmark was identified with two markings per expert. In the localization of each landmark, the experts were asked to consider the spatial location of each landmark and to localize its centroid. To determine the most accurately chosen landmarks we imported all fours sets of landmarks on top of the digitized mandible model and we determined for each landmark the average of its Cartesian coordinates as well as its standard deviation based on the available datasets (i.e. 4 markings total per landmark) Among all pairs of landmarks, of interest was not only their accurate marking but also their palpability given that in the final application we will need to mark those landmarks on a HPS. Among the easily palpable landmarks, landmarks **1** and **7** were found to be the most accurate and thus their distance was chosen as the reference length used in section 4.2. for shape characterization.

Having repeated landmark selection by the same expert and using more than one expert provides us with inter- and intra-expert variability among the two experts in the generation of surrogate of ground truth for landmarks on the mandible [26]. In deciding on the minimum required number of experts needed for a given task, we showed in another application related to a more complex task, that of segmentation of 3D models, that it was sufficient for three experts to repeat delineations three times in order to assess intra- and inter-expert variability for ground truth generation.[27] Hand marking by experts, conducted under the same strict protocol, is still the best method to generate ground truth for the landmarks on the mandible. The choice for only two experts in the marking of landmarks is consistent with the task complexity. A more extensive follow-up study could focus on selecting and digitizing a larger number of mandibles to show variation of human anatomy in different phenotypes, as well as including more experts to further quantify statistical variability.

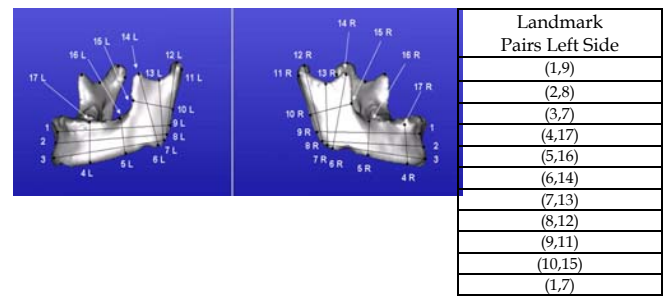

| Landmark Pairs Left Side |
|---|
| (1,9) |
| (2,8) |
| (3,7) |
| (4,17) |
| (5,16) |
| (6,14) |
| (7,13) |
| (8,12) |
| (9,11) |
| (10,15) |
| (1,7) |

Figure 4. Landmark selection and landmark pairs

### 4.2 Mandible Shape Characterization

In order to proceed with the characterization of each mandible, we formed landmark pairs based on the direction of growth of the mandibles [14], obtaining 11 pairs. The landmark pairs are shown in Figure 4. We computed the geodesic distance between the landmarks of each landmark pairs. The distance was computed using the GeoMagic distance tool. Further we devised a shape metric by dividing the distance between the landmark pairs by the distance between landmarks 1 and 7 denoted as R.

We then define two types of ratios:

- The **Local Ratios, denoted as** $LR_{i,j}$, are the ratios of the length $L_{i,j}$ between selected pairs of landmarks $i$ and $j$ to the reference length on the same mandible as illustrated in Figure 6. (i.e. $LR_{i,j} = L_{i,j}/R$)

- The **Global Ratios**, denoted as $GR_{i,j}$, are the ratios of the local ratios for any pair of landmarks across any two mandibles. The hypothesis is that they will be approximately 1 indicating that the local ratios are equal. The $GR_{i,j}$ may be expressed as

$$GR_{i,j} = \frac{LR_{i,j} \text{ for mandible } a}{LR_{i,j} \text{ for mandible } b} \quad (1)$$

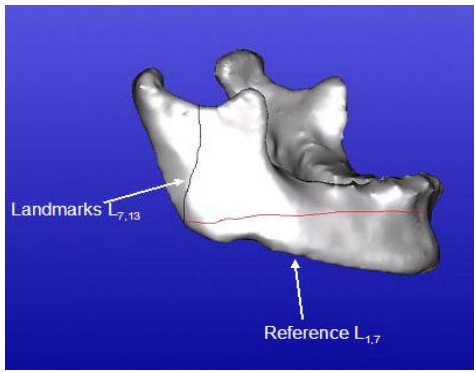

Figure 5. Local Ratio

Given the local ratios for each mandible, the set of global ratios was determined for each pair of mandibles i.e. mandible 1 and 2 (Figure 6a), mandible 2 and 3 (Figure 6b), mandible 3 and 1 Figure (6c). The left and right sides of each mandible were treated independently to capture potential asymmetries and are represented on each figure with a square and a triangle symbol, respectively. The measurements are plotted for each expert separately on the left and right side, respectively.

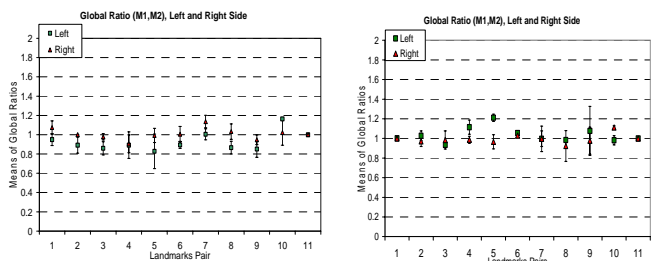

Figure 6a. Global ratios for mandibles 1 & 2, expert 1 on the left, expert 2 on the right

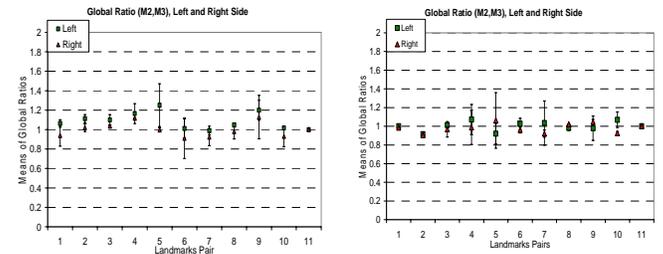

Figure 6b. Global ratios for mandibles 2 & 3, expert 1 on the left, expert 2 on the right

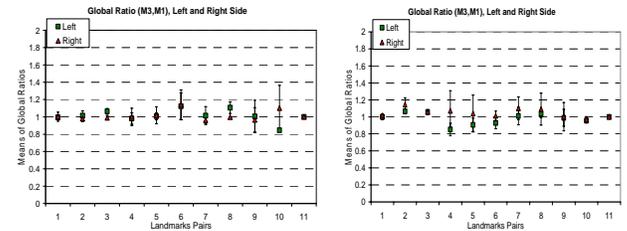

Figure 6c. Global ratios for mandibles 3 & 1, expert 1 on the left, expert 2 on the right

Results show that the global ratios are 1 within a ±20% range, with a RMS variation of 3%. Such results constitute preliminary proof that the distances are proportional which leads us to the following hypothesis: if the distance between two landmarks on the mandible is given, which characterizes the growth under a particular age group, the distance between other landmarks can be computed. Such a relationship allows us to generate scaled 3D mandible models from a given mandible model. The plots in Figure 6 also point us to the symmetry of the right and left sides of the mandibles since the global ratios for the left and right side follow a similar trend.

To quantify the landmark selection we computed the average values of the local ratios over the three mandibles for each pair of landmarks. Figure 7 presents the plots of these averages for each anatomist in each session.

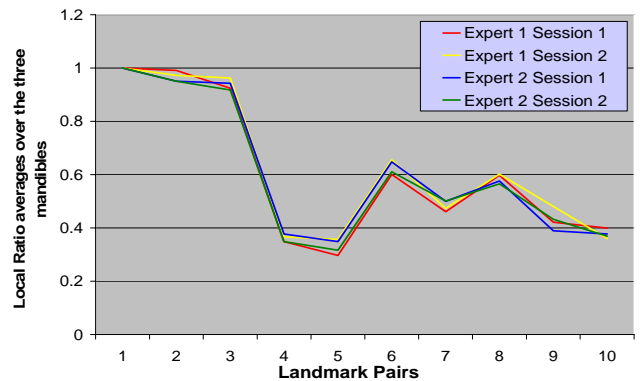

Figure 7. Local Ratio averages over the three mandibles for each expert and session

The plot shows that these values are extremely close to each other, thus pointing to the consistency of the landmark selection process.

### 4.3 Scaling Computation

The method of scaling a generic mandible to a target mandible proceeded as follows. We first marked the positions of the reference landmarks pairs in the given and target mandibles and computed the distance between them along the surface of that mandible. We then computed the ratio of the reference landmarks pair lengths across the two mandibles. This ratio generated a relative scaling factor that was used to scale a given mandible to the target mandible on the directions of growth. From reference [14] figure 4, we established two main direction of growth, one vertical along the chin, and the other along the anterior masseter corner. We computed the precise angle (i.e. 86 degrees) between the two main direction of growth using the scalar product of two unit vectors defined along these two directions using associated landmarks along these directions.

## 5. VIRTUAL MODEL SIMILARITY QUANTIFICATION

To quantify the scaling process we compared one of the generated scaled virtual mandible model to the real counterpart. For this experiment we used two real mandibles denoted M1 and M2. The models currently have a level of similarity with an average departure of 2.15mm mean and 1.32 mm standard deviation. The quantification experiment consisted of the following steps:

1. Obtained a virtual model (VM1) for mandible M1 using the digitization process described in Section 3.1

2. Scaled VM1 using the proposed method in section 4.3 from M1 to M2. Let's call this generated virtual model VM1Scaled.

3. Obtained a virtual model (VM2) for mandible M2 using the same digitization process. This step was necessary only for the purpose of quantification of similarity between the two virtual models VM2 and VM1Scaled.

4. Register the mandible VM2Scaled with VM1 using a rigid body transformation with Geomagic Software. The chin landmarks (1, 2, and 3) of both the mandibles are first associated. The VM2Scaled was then rotated along the vertical chin axis until the distances between the landmarks 7L and 7F of both mandibles were minimized.

5. Compared VM2 and VM1Scaled using RMS error analysis between all the surface nodes of both the mandibles.

In order to perform a preliminary subjective visual assessment we superimposed the two 3D models VM2 and VM1Scaled. The centroid of the models landmarks were matched and the models rotated to match the reference landmarks. One model was set of blue color and the other of grey color. Figure 8 provides a subjective assessment of the high similarity of the two 3D virtual models.

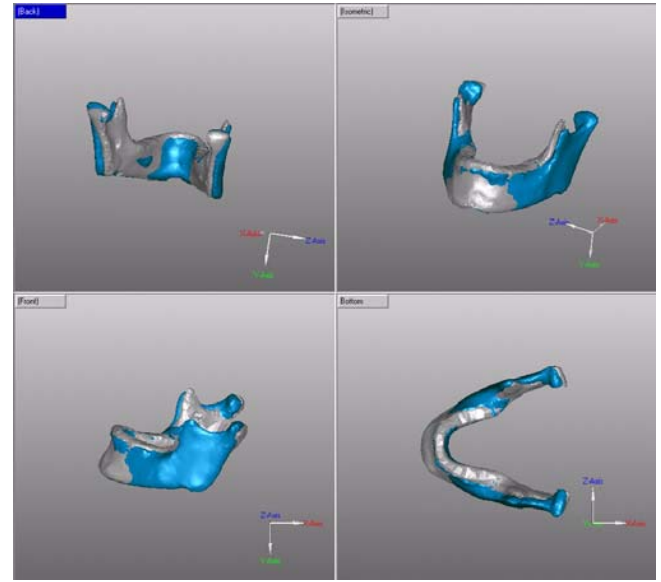

Figure 8. VM1Scaled in Blue color superimposed over VM2 in Grey color

For a quantitative and objective assessment, once the reference landmarks were placed in coincidence, we computed the distance between each pair of vertices in the 3D models and drew a color distance-map that shows the areas of perfect fit (i.e. green color) as well as the largest displacement (i.e. red color) between the vertices of the two models (Figure 9).

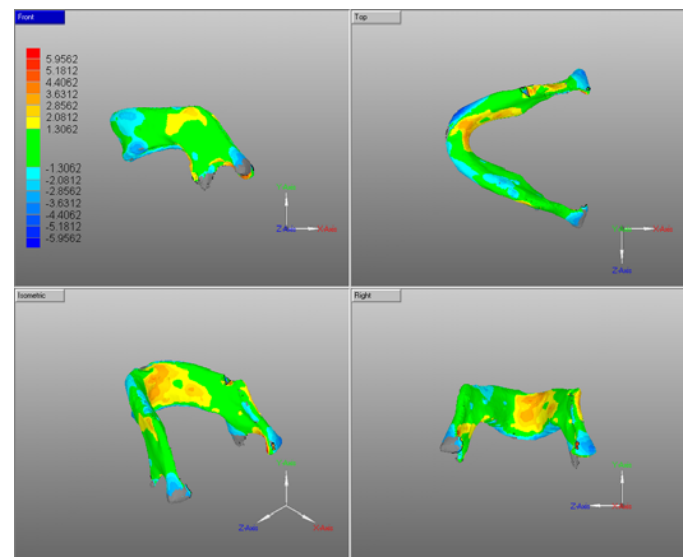

Figure 9. (VM1Scaled ,VM2) Vertices distance map

As we can see in Figure 9, the green color is predominant in the images, which shows a perfect fit between the two virtual models. Table 2 provides a quantitative summary of the results. It is important to note that in the experimental protocol, the fact that a similar digitization process was used in the quantification algorithm allowed us to ignore the digitization error terms since these errors compensated each other when the distance between the pair of vertices was computed.

Table 2. Comparison Statistics

| | |
|---|---|
| Number of polygons per model | 46,950 |
| Number of Vertices per model | 23,478 |
| Maximum distance between two vertices | 5.65 mm |
| Average distance between two vertices | 1.30 mm |
| Standard Deviation | 0.99 mm |

Results show a level of similarity with an average departure of 1.30 mm within a 0.99 mm standard deviation. Such performance well satisfies the requirement of being able to put the models in coincidence within a 3mm Root Mean Square error. Hence the generated model VM1Scaled can be used instead of VM2 eliminating the digitization process and the associated time and resources.

## 6. TOWARDS TEACHING ENDOTRACHEAL INTUBATION

Augmented Reality as a tool for teaching endotracheal intubation was described in [28], where guidance was aimed at "nearly"-correct endotracheal intubation position as shown in Figure 10.

As part of this experiment, MR scans of the larynx of one volunteer were generated as a pilot study, using the Phillips INTERA Gyroscan 1.5T. A sagittal view from head to upper chest was used in order to determine the ideal head and neck position for intubation. Four final MR images, three incorrect (i.e. hyperflexion of the neck, and neutral), and one correct (i.e. "sniffing position," or slight flexion of the neck and extension of the head). In Figure.10 we show two incorrect and one correct intubation positions that were determined visually by the relative patency of the airway passage from oropharynx to trachea.

We suggested how one could potentially guide with the virtual models of the trachea and the mandible (from the Visible Human Male data) training for endotracheal intubation (Figure 11). The Visible Human virtual 3D scene needed then to be rearranged to depict the mandible and trachea in the correct intubation position. The relationship between the Visible Human 3D mandible and trachea was changed from its original position using the angle between the subject mandible and trachea in the MR scan corresponding to the best (yet not necessarily optimal) position. It must be noted that it was nearly impossible to achieve a perfect "sniffing position" looking at the external anatomical landmarks, yet the best "nearly-correct" position was good enough to perform intubation.

Using these methods to rearrange the relationship of the trachea with respect to the mandible, together with the proposed technique to scale the Visible Human mandible to that of the HPS enables visualization of virtual models optically superimposed on the HPS. Knowing what is a nearly optimal relationship between the trachea and the mandible to perform endo-tracheal intubation, it helps to position a mannequin/subject before we will use mandible tracking to calibrate the intubation position even further. The superimposition of a virtual and a properly scaled mandible to the physical one can be used as a didactic presentation in training of intern to perform endo-tracheal intubation.

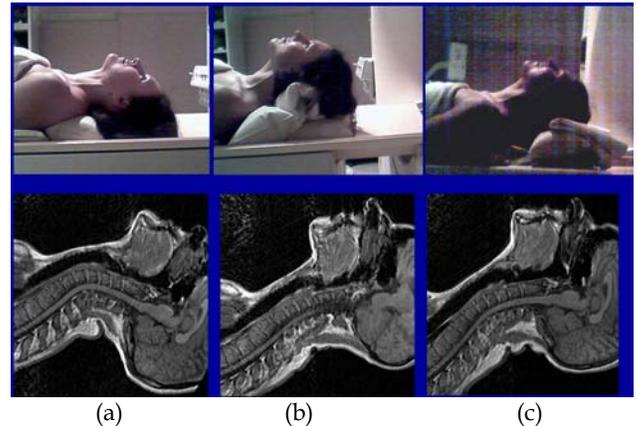

Figure10. Intubation position using external anatomical landmarks: (a)-(b) incorrect and (c) correct intubation positions [28]

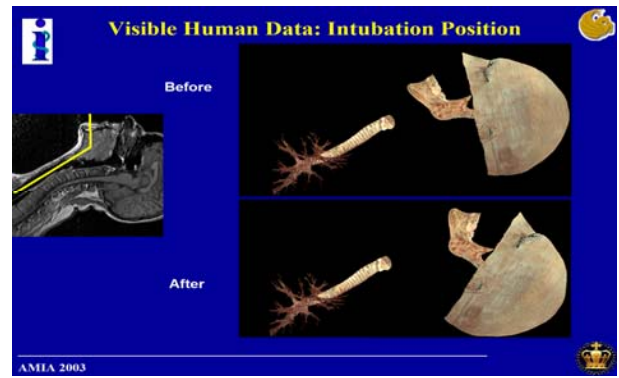

Figure 11. Intubation position in MR sagittal view and using rearranged 3D models derived from the Visible Human dataset [28].

## 7. CONCLUSION AND FUTURE WORK

This paper presents an alternative method to patient specific data acquisition of generating 3D models for Augmented Reality medical applications. We show that 3D anatomically correct mandible models can be scaled considering particular landmarks in order to be made morphometrically equivalent to each other. The method used to compute a scaling factor is simple in implementation, yet it could generate 3D mandible virtual models with vertices within a predicted 1.30 millimeter average error bound from their real counterparts.

We plan, in the future, to investigate a larger number of mandibles that will represent different phenotypes categorized by gender, age, race and other factors in a further validation of the results found in this study. A further validation of the scaling procedure will facilitate applications of bone scaling techniques in different fields. In forensics, for example, parts of the missing pieces of the bone structure might be generated from existing 3D models. Moreover with the increase in the use of digital libraries in the medical field, 3D patient specific data may be collected only once in the patient lifetime and 3D models of his/her bone structure might be regenerated at any time during the course of the patient life without the need of rescanning, by scaling the existing models using a specific landmarks based on the bone growth patterns.

## 8. ACKNOWLEDGEMENTS

We wish to thank our sponsors the NSF/ITR IIS-00-820-16, the Link Foundation, the Office of Naval Research Grant N000140310677, the US Army Simulation, Training, and Instrumentation Command (STRICOM), and the Florida Photonics Center of Excellence for their invaluable support for this research. We also thank METI Corporation for providing us through the US Army with the upper torso of the Human patient simulator and Karen Kerner for the intubation position images using external anatomical landmarks.

**Neha R. Hippalgaonkar and Alexa D. Sider** are high school students at Lake Highland Preparatory School. They are involved in interdisciplinary research at Optical Diagnostics and Applications Laboratory at the School of Optics at University of Central Florida. Their interest is virtual models generation for Augmented Reality environments.

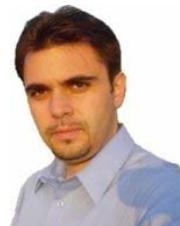

**Felix G. Hamza-Lup, Ph.D** is Assistant Professor of Computer Science at Armstrong Atlantic State University, Savannah. He received his B.Sc. in Computer Science from the Technical University of Cluj-Napoca, Romania his M.S and Ph.D. in Computer Science from the University of Central Florida. His current focus is the development of novel distributed applications based on Augmented Reality paradigms for the medical field, visualization tools and algorithms that enhance consistency in distributed interactive virtual environments. Felix has been awarded the Link Foundation Fellowship in Advanced Simulation and Training and the Hillman Award for distinguished Doctoral Research in Computer Science. His research interests include, distributed systems and applications, virtual and mixed reality environments, human computer interaction and motion tracking sensors. He is a member of the ACM, IEEE, SPIE and the International Honor Society for the Computing Sciences.

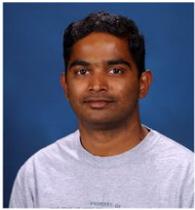

**Anand P. Santhanam, M.S.** is a doctoral candidate at the School of Computer Science at the University of Central Florida. He received his B.E degree in Computer Science from the University of Madras, India, in 1999, and his M.S degree in computer science from the University of Texas, Dallas in 2001. He worked as a software engineer at Nortel Networks, Richardson TX and at Metera Networks, Richardson TX for the development of network models and routing protocols. His current research interests include real-time graphics and animation in networked virtual environments. Anand has been awarded the Link Foundation Fellowship in Advanced Simulation and Training and the Graduate Research Excellence Award in UCF.

**Gerald Bertetta,** is an anatomist and instructor of physical therapy at the College of Health and Public Affairs at the University of Central Florida. His research interests include anatomical models and forensics.

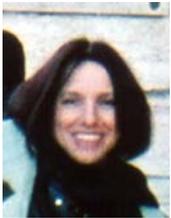

**Celina Imielińska, Ph.D.** is an electrical engineer and computer scientist, and an Associate Research Scientist affiliated with Columbia University College of Physicians and Surgeons Office Scholarly Resources and Department of Medical Informatics, and Department of Computer Science. She has M.E. degree in electrical engineering from Politechnika Gdanska, in Gdansk, Poland; and M.S. and Ph.D. in Computer Science from Rutgers University, in New Brunswick, NJ. Her current interests are medical imaging, image segmentation, 3D visualization, and computational geometry.

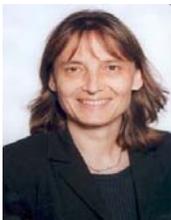

**Jannick P. Rolland, Ph.D** is Associate Professor of Optics, Computer Science, Electrical Engineering, and Modeling and Simulation at the University of Central Florida. She received a Diploma from the Ecole Superieure D'Optique in Orsay, France, in 1984, and her Ph.D. in Optical Science from the University of Arizona in 1990. Jannick Rolland then joined the Department of Computer Science at the University of North Carolina at Chapel Hill (UNC-CH) as a Postdoctoral Student to conduct research in optical design for 3D medical visualization. She was appointed Research Faculty at UNC in 1992 and headed the Vision Research Group from 1992 to 1996. She holds five patents, wrote 6 book chapters, and has over 50 peered review publications related to optical design, augmented reality, vision, and image quality assessment for medical imaging. Dr. Rolland is Associate Editor of Presence (MIT Press), and has been Associated Editor of Optical Engineering 1999-2004. She is the UCF Distinguished Professor of year 2001 for the UCF Centers and Institutes, a member of IEEE, and a Fellow of the OSA.